# Simple metal and binary alloy phases based on the hcp structure: electronic origin of distortions and superlattices

**Valentina F Degtyareva and Nataliya S Afonikova**

Institute of Solid State Physics, Russian Academy of Sciences,
Chernogolovka 142432, Russia
E-mail: degtyar@issp.ac.ru

**Abstract.** Crystal structures of simple metals and binary alloy phases based on the close-packed hexagonal (*hcp*) structure are analyzed within the model of Fermi sphere – Brillouin zone interactions to understand distortions and superlattices. Examination of the Brillouin-Jones configuration in relation to the nearly-free electron Fermi sphere for several representative phases reveals significance of the electron energy contribution to the phase stability. This approach may be useful for understanding high pressure structures recently found in compressed simple alkali and alkali-earth metals.

## 1. Introduction

Crystal structures of metals are based on the close-packed symmetrical types: face-centered cubic (*fcc*), close-packed hexagonal (*hcp*) and body-centered cubic (*bcc*) structures. Formation of these structures is governed by the electrostatic (Madelung) energy as the main contribution to the crystal energy. A significant contribution gives the band structure energy term related to the valence electron energy that reduced by formation of the energy gap on the Brillouin zone (BZ) planes near the Fermi sphere (FS) [1,2]. The latter effect is considered within the nearly-free electron model and called the FSBZ interaction. This model has been developed to explain the structural sequence that occurs in the family of Hume-Rothery phases in the Cu-Zn and related alloys along the alloy composition [3]. Hume-Rothery effect has been identified to play a role in the stability of a wide variety of materials, such as structurally complex alloy phases, quasicrystals and their approximants [4]. Liquid and amorphous metals have also been considered to obey the Hume-Rothery rule [5,6].

Within the nearly-free electron model the energy gap is formed on a Brillouin zone plane when the Fermi sphere radius, $k_F$, equals to the half of wave vector $q_{hkl}$ of that plane, so that $k_F = \frac{1}{2} q_{hkl}$. The symmetrical close-packed structures *fcc*, *bcc* and *hcp*, together with the complex cubic γ-phase $Cu_5Zn_8$-*cI*52, form a basic Hume-Rothery phase sequence in which the stability of each phase is defined by the number of valence electron per atom, or electron concentration, z. In binary alloys there is vast variety of structures arising due to importance of some other factors as atomic size difference, electronegativity etc. Nevertheless, stoihiometric compounds are stabilized within the alloy composition defined by effect of FSBZ interaction.

Binary phases of simple metals form an area of exploration of physical-chemical factors of structural stability. The interest in this problem has been renewed because of the recent discoveries of complex structures in elemental metals under pressure (see reviews [7-10] and references therein). These relations of the elemental structures to binary phases imply some similarities in the physical origin of structural stability. The stability of the complex structures of alkali metals on pressure increase can be related to the Hume-Rothery mechanism assuming an electron transition from core-to-valence band as suggested for sodium in the *oP*8 structure [11]. This phase is structurally identical to some of the binary phases, as for example the AuGa-*oP*8 with 2 electrons per atom [12]. Therefore an analysis of structural relations for binary phases and mechanisms of the distortion from high symmetry metallic phases may help to understand the origin of complexity in compressed elements.

In our recent paper we considered structures of simple metals based on the *bcc* structure with distortions, vacancies and superstructures [13]. The subject of the present paper is *hcp* and derivative structures in simple metals and binary phases formed by *sp* elements. Consideration of *sp* (non-transition) metals gives well-defined value of the valence electron count per atom in a phase to account for the Fermi sphere radius and to compare it with the BZ configuration.

For the classical Hume-Rothery phases the matching rule is a good accommodation of the FS by the respective Brillouin zone boundaries with the strong structure factor. As was considered by Heine and Weair [14] "for polyvalent metals there are always many zone planes cutting the Fermi surface, and the contributions to the intersection of two or more zone planes may in some cases lead to important effects"; therefore "one is tempted to consider the Jones zone as approximated by a sphere". To meet this requirement experimentally observed tetragonal distorted *fcc* structures in In and In alloys has been found to follow alloy electron concentration [15,10]. It was shown that the degree of the *fcc* distortion is defined by accommodation BZ within FS and a contact BZ corners with FS. Similar cases are considered in the present paper as factors for distortion of *hcp* with variations of the axial *c/a* ratio.

Some representative examples of complex binary phases are considered that can be derived from the *hcp* structure. Alloy phase components are selected from simple metallic elements (Cu and Au with addition of neighboring polyvalent metals), so that the valence electron contribution of constituents is defined by the number of the group in the Periodic Table.

**2. Method of analysis**

Crystal energy contribution due to the band structure energy can be estimated by configurations of Brillouin–Jones zone planes in the nearly free-electron model. A special program BRIZ has been developed to construct FS-BZ configurations and to estimate some parameters such as the Fermi sphere radius ($k_F$), values of reciprocal wave vectors of BZ planes ($q_{hkl}$) and volumes of BZ and FS [16]. The BZ planes are selected to match the condition $q_{hkl} \approx 2k_F$ that have a non-zero structure factor. In this case an energy gap is opened on the BZ plane leading to the lowering of the electron energy. The ratio of ½$q_{hkl}$ to $k_F$ is usually less than 1 and equals ~0.95, called as "truncation" factor [17]. In the FS-BZ presentations by the BRIZ program the BZ planes cross the FS, whereas in the real system the Fermi sphere is deformed (remaining inside BZ). The "truncation" factor has a characteristic value and corresponds to a decrease in the electron energy on the BZ plane.

The crystal structure of a phase chosen for the analysis by the BRIZ program is characterized by the lattice parameters and the number of atoms in the unit cell, which define the average atomic volume ($V_{at}$). The valence electron concentration (z) is the average number of valence electrons per atom that gives the value of the Fermi sphere radius $k_F = (3\pi^2 z / V_{at})^{1/3}$. Further structure characterization parameters are the number of BZ planes that are in contact with the FS, the degree of "truncation" factor and the value of BZ filling by electronic states, defined as a ratio of the volumes of FS and BZ. It should be noted that in the current model the FS volume is a measure of the number of valence electrons participating in the band structure contribution even when the FS is deformed from a sphere in the real case. As discussed by Heine [14], the Brillouin-Jones zone – also called as "Jones zone" (or "Jones barrel") – should accommodate well a FS; consequently the special stability of metallic structures is associated with the condition of "Jones zone being completely filled" with electronic states. As was discussed in Introduction, for polyvalent metals with large FS it is important to consider configurations of the FS and the BZ that are accommodated by the FS.

Presentations of the FS-BZ configurations are given with the orthogonal axes with directions in the common view when a* is looking forward, b* - to the right and c* - upward. For the hexagonal lattice in reciprocal space a*= $a_{1h}$*cos30°, b*= $a_{2h}$* and c*= $c_h$*. Some FS-BZ views are given along c*. All structural data for binary phases considered in this paper have been found in the Pauling File [18] and the standardized crystallographic data are used.

## 3. Results and discussion

### 3.1. Axial c/a distortions of the hcp structure

Binary alloys of the noble metals with the following B Group metals form three independent phases with the *hcp* structure: ζ, ε and η as discussed by Massalski and King [19] and by Pearson (Ref. [12] pp.120-121).These phases show a characteristic variation of the axial ratio with the electron concentration per atom deviating from the ideal value c/a=$\sqrt{8/3}$ =1.633. We consider *c/a* variations in relation to FS-BZ configurations. Estimations within the nearly-free electron model give for the contact of FS with the BZ planes (100), (002) and (101) the values of electron concentration 1.14, 1.36 and 1.65, respectively. Assuming the "truncation" factor ~1.05, these values should be extended by factor ~$1.05^3$ to 1.33, 1.58 and 1.9. The *hcp* phase with the axial ratio $c/a \approx 1.65$, called as ζ, occurs at $z \approx 1.2$–1.4 and with increasing of electron concentration a decreasing *c/a* is observed in the region of ε-phases until at $z \approx 1.85$ where *c/a* starts increasing rapidly in the region of η-phases.

#### 3.1.1. The ζ–hcp phase with c/a ~ 1.65

The *hcp* phase with *c/a* ~ 1.65 is represented by Au₃Hg (see Table 1 and Fig.1a). The FS is located within the hexagonal prism formed by planes (100) and (002). There is slight overlap of the ideal FS with (100) planes and $k_F$ is below the (002) planes. With this configuration one should expect an attraction of FS and (002) planes, that should produce increasing of *c/a* comparing with the ideal value.

The region of $z \approx 1$–1.4 in the Cu-Zn system is occupied by the *fcc* structure, however for the Au-based alloys there is a competition between *fcc* and *hcp* because heavier elements are more responsive to the band structure energy term. Thus in pure elements Cu and Ag *fcc* is found to be stable up to 190 GPa and 92 GPa, respectively [8] and for Au the *fcc* – *hcp* transition was found above 250 GPa with nearly ideal axial ratio *c/a* = 1.63 [20].

At z >1.4 the axial ratio decreases down to $c/a \approx 1.55$ caused by FS that "attracted" (101) BZ planes and "repulsed" (002) BZ planes resulting in extension and contraction of lattice parameters *a* and *c*, respectively, and the decrease of *c/a*.

#### 3.1.2. The classical hcp Hume-Rothery phase Cu₂₀Zn₈₀

The $Cu_{20}Zn_{80}$ phase is one of the representative Hume-Rothery phases, called as ε, occurring at electron concentration z ~ 1.8. Structural data and parameters of the FS-BZ configuration are given in Table 1 and presented in Fig.1b. The free-electron sphere is in contact with the BZ planes (002) and (101) with truncation factors 1.068 and 1.033, respectively. The degree of BZ filling by electron states is ~87% which satisfies the Hume-Rothery matching rule. It should be noted that the FS accommodates well the hexagon polygons formed by planes (100), as shown in Fig.1b right.

Same phases, called ε, exist in related binary systems such as Cu-Sb, Ag-Cd, Ag-In, Ag-Sb (see Ref. [12] pp. 116-126). In the systems Au-Ge and Ag-Ge with the phase diagram of the simple eutectic type the *hcp* phases were found by quenching from the melt [21] and after pressure action [22,23] in the range of electron concentration z = 1.5 – 1.8 corresponding to 16 – 27 at.% Ge.

To this family of *hcp* phases also belongs the *hcp* phase found in Ba under pressure above 5.5 GPa (Ba-*hcp*I), where outer *s* electrons are expected to transfer partially to the higher empty *d*-band [24,25]. The axial ratio *c/a* for Ba-*hcp*I is 1.55 at 5.9 GPa and it was found to decrease with pressure approaching the value of *c/a* = 1.50 at 12.6 GPa which is an indication of the continuous process of the *s – d* electron transfer. Another *hcp* phase in Ba (Ba-*hcp*II) is found to occur at pressures above 45 GPa and obviously belongs to another family of the *hcp* phases, which is related to *hcp* in Pb and other group IV elements (see in section 3.2).

*3.1.3. Phases hcp in divalent metals Zn and Cd*

The next region of the *hcp* phase, called as η, is formed in the pure elements Zn and Cd and in solid solutions based on these elements. The *c/a* axial ratios for Zn and Cd are 1.856 and 1.886, respectively which is considerably higher than the ideal *c/a* value. The *c/a* ratio for a divalent metal can be estimated assuming the contact of the FS with the corners of the hexagon formed by planes (100). This condition corresponds to the relation $k_F = ⅓ q_{110}$ giving *c/a* = 1.86 which is close to the experimental value of *c/a* for Zn (see Table 1 and Fig.1c, middle). For the ideal value *c/a* = 1.633 there is no contact of the FS with the corners of the accommodated hexagon (Fig. 1c, right) and no gain in the band structure energy due to this reason.

In alloys of Cd with Hg the *hcp* phase is formed at composition up ~25 at% Hg with the ratio *c/a* ≈ 1.91 which decreases monotonically to ~1.65 under pressure to 50 GPa [26,27]. In alloys of Zn with Hg the *hcp* phase formed at composition ~70 at% Zn has the ratio *c/a* ≈ 2.04 [28]. In mercury under normal pressure a rhombohedral structure (distorted *fcc*) is stable and the *hcp* phase appears at pressure above 40 GPa [29] with *c/a* ≈ 1.75, the Hg-*hcp* remains stable up to ~190 GPa [30].

The *hcp* phases in Zn and Cd are stable at least to 126 GPa and 174 GPa, respectively, with decreasing of *c/a* on compression [31,32]. The *c/a* axial ratios continuously decrease with pressure down to a value of 1.59. A clear change in the slope of the volume dependence of the axial ratio at $c/a = \sqrt{3}$ has been observed for both metals. The anomaly may be related to the electronic topological transition, as discussed in [33]. Within the model of FSBZ interactions the *c/a* behavior for Zn and Cd under pressure is related to the balance between two energy contributions – electrostatic and band structure – that scale different with volume.

*3.2. The hcp structure in polyvalent group IV elements*

The *hcp* structure can be stabilized in elements at higher valence electron counts than considered above. Group IV elements transform to *hcp* under high pressure: Si at 42 GPa, Ge at 160 GPa, Sn at 157 GPa, and Pb at 13 GPa [7,8,34]. Structural data for Pb-*hcp* at ~14 GPa are given in Table 1 and the BZ construction is shown in Fig.1e. For the valence electron number z = 4 the position of $2k_F$ is close to the diffraction peak (102), and FS is in contact with the BZ planes of (102) type. The axial ratio for Pb-*hcp* is *c/a* ≈ 1.65 and remains nearly stable in the wide pressure range from 13 to 110 GPa, showing only small changes on compression [35,36]

The *hcp* phase was found to form in binary Sn-based alloys with In and Hg at compositions corresponding to ~3.8 electrons/atom at pressures above 15 GPa and 50 GPa, respectively [37,38] which is significantly less than for pure *hcp*-Sn. These phases may be considered as extended solid solutions of In and Hg in *hcp*-Sn. In similar way may be considered the *hcp* phase in $Al_{30}Ge_{70}$ formed at pressures above 50 GPa [39]. There is a good contact of the BZ planes (102) and the Fermi sphere with the "truncation" factor of ~1.027 (for z = 3.8) or ~1.018 (for z = 3.7) that is slightly less than the value $k_F/(½ q_{102}) = 1.045$ for *hcp*-Pb (see Table 1, last column). The *hcp*-$Pb_{80}Bi_{20}$ phase exists at ambient pressure [18] and stabilized by same mechanism with $k_F/(½ q_{102}) = 1.061$, whereas the *hcp*-$Pb_{80}Sb_{20}$ is formed under pressure [40].

It is important that the FS accommodates well the polyhedron formed by (101) planes related to the first strong diffraction peak (see Figure 1e, right). This is in agreement with the importance of the case when "the Jones zone is approximated by a sphere" [14], as was discussed in Introduction. Same reason can be identified for the stability of the *hcp* structure in the group III element Tl with 3 valence electrons at ambient conditions with *c/a* = 1.60. Under pressure *hcp*-Tl transforms to *fcc* at 3.7 GPa [41]. The lightest group III metal aluminum was found to transform from *fcc* to *hcp* at pressure ~220 GPa with the nearly ideal *c/a* ratio (~1.62, which is pressure independent) [42]. The high pressure *hcp*-Al has a resemblance to the ambient pressure *hcp*-Tl, and these 3-valent metals both may have common energetical origin for stabilization of the *hcp* structure although in a different pressure range.

To this group of *hcp* phases stabilizing at z = 4 is related the Ba-*hcp*II found at 45-105 GPa [24,25] with the nearly constant *c/a* (1.572 at 53 GPa ). The Ba-*hcp*I occurred at 5.5-12 GPa is characterized by significant decrease of *c/a* on compression and is related to *hcp* phases similar to that in Cu-Zn alloys, considered in 3.1.2. The *c/a* ratio is less than the ideal value of 1.633 and approaches 1.50 at 12 GPa. The valence electron count in Ba-*hcp*I is defined by the *s – d* transfer giving z less than 2.

For Ba-*hcp*II it is necessary to assume the overlap of the valence electron band and upper core levels giving z value approximately 4 as for group IV elements. Taking lattice parameters of Ba-*hcp*II at 53 GPa *a* = 3.1035 Å, *c* = 4.8778 Å [24] one obtains volume compression $V/V_0$ = 0.32 and the atomic radius $r_{at}$ = 1.55 Å which is significantly less than 1.61 Å – the ionic radius $Ba^{2+}$ estimated by Shannon [43]. Thus, at such strong compression ionic radii of Ba should overlap leading to repulsion, and for stability of the structure it is necessary to assume an electron transfer from the outer core to the valence band. This core ionization will lead to increasing the valence electron count to four as in the case of group-IV elements with the *hcp* structure.

The valence band – core level overleap was considered previously for Na in the high pressure *oP*8 structure [11]. Valence electron count equal to four is suggested for alkali metals K, Rb and Cs in the high pressure structure *oC*16-*Cmca* which is similar to that in four-valent metals Si and Ge [44]. For lighter alkali-earth metal Ca compressed above 30 GPa a transition to the simple cubic structure has been observed that can be explained by a core – valence band electron transfer with increase a valence electron counts to ~3.5 [45].

### 3.3. Superlattices based on the hcp structure

#### 3.3.1. Double hexagonal close-packed structure $Au_{87}In_{13}$

Formation of superstructures based on the close-packed hexagonal cell appears in the case of a double hexagonal close-packed structure with doubling of c axis and with the sequence of close-packed layers ABACA. The structure is called further *dhcp* or *hP*4 (in Pearson notation). Example of this phase is given for $Au_{87}In_{13}$ (Table 1, Fig. 1d). There is no ordering of elements in this phase, and energtical origin for superstructure formation is based on formation of the additional BZ planes (101) type close to the FS. Same *dhcp* phases were found in alloys $Au_{87}Ga_{13}$ and $Au_{71}Cd_{19}$ [18].

#### 3.3.2. Long-period superlattices based on hcp $Au_2Cd$-*hP*98

Another type of superstructures based on *hcp* is long-period superlattices along the *a* axis. An example of this type of superstructure is $Au_2Cd$-*hP*98 with the sell parameters a = $7a_0$ and c = $c_0$, were $a_0$ and $c_0$ are referred to the basic *hcp* cell [46]. Structural data for $Au_2Cd$-*hP*98 and evaluation data with the BRIZ program are given in Table 2. Figure 2a shows BZ construction. It is remarkable that for this structure the BZ consist of the hexagonal prism with (700) planes as for usual *hcp* and addition 6 planes of (440) type giving the 12-fold sided prism. This modification of the BZ corresponds to the gaining in electronic energy within the Hume-Rothery mechanism. How long should be superlattice is defined by the closeness of the superstructure reflections to $2k_F$: ½ $q_{440}$ ≈ ½ $q_{700}$ ≈ $k_F$.

Similar type of the *hcp* superstructure was found in the $Au_{11}In_3$ phase [47,48]. An approximant structural model of $Au_{11}In_3$ is suggested with cell parameters a = $15\sqrt{3} a_0$ and c = $c_0$. The strongest modulation reflections of (26.0.0) type have vector length very close to the (15.15.0) type reflections, or (100) type for the basic *hcp*: $q_{26.0.0} ≈ q_{15.15.0}$. Constructed BZ configuration would be similar to that for $Au_2Cd$-*hP*98.

From the FS-BZ configuration for $Au_2Cd$-*hP*98 in projection along c* (Fig.2a, right) one can see that the 12-sided polygon is accommodated well by the FS. That leads to additional electron energy gaining and should be considered as origin of superlattice formation. The regular 12-sided polygon is formed by BZ planes if the superlattice period is defined by $\sqrt{3}/\tan 15°$ (=6.4641), as was estimated by consideration of the incommensurate superstructure $Au_{11}In_3$ based on the *hcp* structure [47,48]. This phase corresponds to the electron concentration value z = 1.43 giving $k_F$ accommodating well the 12-sided polygon (dodecagon) with small exceed of $k_F$ above the corner of polygon (at ~1.04) satisfying the Hume-Rothery rule.

*3.4. Orthorhombic distortions of the hcp structure*

*3.4.1. The structure $Cu_3Sn$-oC80*

Superstructure $Cu_3Sn$-*oC*80 is based on the orthorhombic subcell related the *hcp* with the parameters a = 2$a_0$, b = 10$b_0$, c = $c_0$. Close packed layers are parallel to the (001) plane and they are stacked along [001] with the size of antiphase domains ~5$b_0$ [49]. As a result of formation of this long-period superlattice new Brillouin zone planes are arising close to the FS (see Table 2 and Fig.2b). A series of planes ((002), (022), (062), (082)) accommodates the FS producing a reduction of electron energy.

It is remarkable that the superstructure period (b = 10 $b_0$) for $Cu_3Sn$ is the same as for the classical CuAu alloy with the face-centered tetragonal basic cell [17,50]. Thus, the origin of superstructure formation has common features and is irrespective of structure.

*3.4.2. The $Cu_3Ge$-oP8 phase*

Orthorhombic $Cu_3Ge$-*oP*8 has atomic arrangement in close-packed layers parallel to the (010) plane and cell parameters related to the *hcp* a = $c_h$, b = 2$a_h$, c = $a_h\sqrt{3}$. This structure is of ordered type defined by an alloy composition and exists in some other compounds such as $Ag_3Sn$, $Cu_3Sb$, $Au_3In$. This type of structure was found in the Ag-Bi alloy subjected to ~7GPa [51]. FS-BZ configurations for $Cu_3Ge$-*oP*8 (Fig.2c) show formation of BZ planes (012) and (201) in addition to the basic BZ planes characterized the *hcp* structure.

*3.4.3. The phases $Cu_{10}Sb_3$-hP26 and $Cu_{11}Sb_3$-oC28*

The phase $Cu_{10}Sb_3$-*hP*26 is characterized by atomic positions formed by slightly distorted close packed layers at high z = ¼ and ¾ with atomic order. The $Cu_{10}Sb_3$ stoichiometry is satisfied in each layer and new cell axes are related to the basic *hcp* cell as **a** = 4$\mathbf{a_0}$ + $\mathbf{b_0}$, **b** = 3$\mathbf{b_0}$ - $\mathbf{a_0}$. The FS-BZ configuration of $Cu_{10}Sb_3$-*hP*26 is shown on Fig.2d with the (131) and (002) basic planes (left) and with additional (311), (400) and (112) planes (right).

Estimations of the FS-BZ configuration for $Cu_{10}Sb_3$ (Table 2) by assuming the valence electron number for Sb z = 5 give the volume of FS grater than the volume of BZ. It is necessary to take into account the difference in the atomic size for Cu and Sb (1.28 Å and 1.59 Å, respectively (see Ref. [12], p.151). Due to size difference of about 25% there are emerged effects of "chemical" pressure in the formation of Cu-Sb alloys, resulting in some changes of valence electron configuration for Sb. As was discussed by Pearson (see Ref.[12], p.231), the higher empty *d*-states may participate with the s and p valence states of group V elements and this transference "can be accounted for the degree of filling of bond orbitals by valence electrons". Within our model only *sp* electron are considered for the FS formation, and the effective number of valence electrons for Sb is assumed as 4, giving for $Cu_{10}Sb_3$ the average number z ~ 1.7. For the next phase with the structure $Cu_{11}Sb_3$-*oC*28 same procedure leads to assume z(Sb) = 4.5 and effective z = 1.75 (see Table 2).

The phase $Cu_{11}Sb_3$-*oC*28 has orthorhombic structure derived from *hcp* as a result of ordering and superstructure formation with cell parameters a = $c_h$, b = 7$a_h$, c = $a_h\sqrt{3}$. The FS-BZ configuration for $Cu_{11}Sb_3$ consists of the basic *hcp* planes (200), (102), (171) and additional planes arising from distortion and superlattice, as planes (180) type shown on Fig. 2e.

In the system Cu-Sb at composition ~20 at.% Sb several long-period modulated phases based on *hcp* have been observed [52-54]. These phases are considered consisting of 7$a_0$ structure and domain walls forming superstructures with the incommensurate period which depends on composition and temperature of annealing. There is a relation of long-period superlattices in the $Cu_5Sb$ alloys to the $Au_2Cd$ phases of $hP98$ type with 7×7 superlattice of $hcp$ considered above. Additional superlattice reflections by formation of the $7a_0$ superstructure result in appearing of additional BZ planes (Fig. 2a) responsible for stabilization of the structure. Important is the configuration of the regular 12-sided prism that accommodates the FS or is inscribed by the FS depending on the alloy valence electron concentration. At $z = 1.6$–$1.8$ electrons/atom there should be FS in contact with the edges of intersections of the planes (101) and (100), as in the case of the $Cu_5Sb$ $hcp$-based structure.

Recently found incommensurately modulated structure of the $LiZn_{4-x}$ ($x = 0.825$) compound is a derivative from the $hcp$ structure with the approximant structure $oC28$ [55]. This structure is closely related to the phase $Cu_{11}Sb_3$-$oC28$ and is stabilized at $z = 1.76$ as is defined for $Cu_{11}Sb_3$ (see Fig. 2e). Authors [55] indicated for $LiZn_{4-x}$ the free-electron-like behavior, and this conclusion gives promise to explain the structural modulations within the Hume-Rothery mechanism.

## 4. Conclusion

Crystal structures based on the close hexagonal packing are considered for simple metals and binary alloy phases to focus attention on the electronic origin of distortions and superlattices. Construction of the Brillouin-Jones polyhedra is done by using the planes corresponding to the diffraction peaks that have wave vectors close to $2k_F$ have demonstrated the Hume-Rothery mechanism to be the main effect that controls phase stability. In such constructions, there are many BZ planes in contact with the FS which gives a high value of BZ filling by electronic states. The variation of the axial ratio $c/a$ is accounted for by the interaction of the Fermi sphere with the basic Brillouin zone planes (100), (002) and (101).

The $hcp$ structures are found in the wide region of valence electron counts for metals from main groups I to IV (including ambient and high pressure phases) and in the alloys of these metals. For polyvalent metals such as the group IV elements the crucial factor is the FS contact with the (102) planes. In these cases it is also important how FS accommodates BZ formed by first strong reflections of (101) type – the contact of FS to the corners of this (101) polyhedron leads obviously to the optimal energy gain. Simple arguments of stability of the $hcp$ structure for divalent elements Zn and Cd with the high value of $c/a$ ratio are found from the tuned alignment of the corners of the (100) polygon to be accommodated by the FS.

The suggested model of the FSBZ interaction helps us establish the mechanism and find the physical origin of the formation of complex superstructures based on a close hexagonal packing. This approach will be useful in understanding formation and structural relationship of structural sequences recently found in elements under high pressure. Unlike in a binary phase, no atomic differences occur in a pure element and thus factors of electronic band structure energy become more apparent especially as they increase under pressure. It is remarkable that $hcp$ structures were found in Ba in two pressure regions with different dependence of $c/a$ on compression that should be accounted for different valence electron states in these phases. The great variety of complex structures in alkali and alkali-earth metals under pressure needs further understanding, in particular their physical origin and relation to basic metallic structures.

**Acknowledgments**

The authors gratefully acknowledge Dr. Olga Degtyareva for valuable discussion and comments. This work is supported by the Program "The Matter under High Energy Density" of the Russian Academy of Sciences.

**Table 1.** Structure parameters of several representative Hume-Rothery phases in binary systems based on group-I elements. Pearson symbol, space group and lattice parameters are from literature data. Fermi sphere radius $k_F$, ratios of $k_F$ to distances of Brillouin zone planes $½q_{hkl}$ and the filling degree of Brillouin zones by electron states $V_{FS}/V_{BZ}$ are calculated by the program BRIZ.

| Phase | $Au_3Hg$ | $Cu_{20}Zn_{80}$ | Zn | $Au_{87}In_{13}$ | Pb (HP) |
|---|---|---|---|---|---|
| *Structural data* | | | | | |
| Structure type, | *hcp* | *hcp* | *hcp* | *dhcp* | *hcp* |
| Pearson symbol | *hP2* | *hP2* | *hP2* | *hP4* | *hP2* |
| Space group | *P6₃/mmc* | *P6₃/mmc* | *P6₃/mmc* | *P6₃/mmc* | *P6₃/mmc* |
| Lattice parameters (Å), $c/a$ | $a$ = 2.918<br>$c$ = 4.811<br>1.649 | $a$ = 2.738<br>$c$ = 4.294<br>1.568 | $a$ = 2.665<br>$c$ = 4.947<br>1.8562 | $a$ = 2.909<br>$c$ = 9.500<br>3.2657<br>$c/2a$ = 1.633 | $a$ = 3.265<br>$c$ = 5.387<br>1.650 |
| Same phases | Au - Cd<br>Au - In | $Ag_{20}Cd_{80}$<br>Ag-In(25-40%In)<br>Ag-Sn(12-23%Sn)<br>Ag-Sb(10-16%Sb)<br>Ba (HP-I) | Cd<br>$Cd_{80}Hg_{20}$<br>Hg (HP) | $Au_{87}Ga_{13}$<br>$Au_{71}Cd_{19}$<br>Cs (HP) | Si (HP)<br>Ge (HP)<br>Sn (HP)<br>Ba (HP-II)<br>In-Sn (HP)<br>Hg-Sn (HP)<br>Pb-Bi<br>Pb-Sb (HP) |
| *FS – BZ data from the BRIZ program* | | | | | |
| z (number of valence electrons per atom) | 1.25 | 1.8 | 2 | 1.26 | 4 |
| $k_F$ (Å$^{-1}$) | 1.278 | 1.561 | 1.573 | 1.289 | 1.683 |
| Total number BZ planes | 8 | 14 | 8 | 32 | 18 |
| $k_F/(½q_{hkl})$<br>*hkl*, max<br>*hkl*, min | *100*: 1.028<br>*002*: 0.979 | *002*: 1.068<br>*101*: 1.033 | *002*: 1.238<br>*101*: 1.047 | *100*: 1.034<br>*102*: 0.913 | *102*: 1.045<br>*110*: 0.875 |
| $V_{FS}/V_{BZ}$ | 0.625 | 0.873 | 1.0 | 0.751 | 0.785 |

Notes: structural data at ambient pressure from [18], data at high pressure (HP) from papers cited in the related paragraphs of the text.

**Table 2.** Structure parameters of several representative Hume-Rothery phases in binary systems based on group-I elements. Pearson symbol, space group and lattice parameters are from literature data. Fermi sphere radius $k_F$, ratios of $k_F$ to distances of Brillouin zone planes $½q_{hkl}$ and the filling degree of Brillouin zones by electron states $V_{FS}/V_{BZ}$ are calculated by the program BRIZ.

| Phase | $Au_2Cd$ | $Cu_3Sn$ | $Cu_3Ge$ | $Cu_{10}Sb_3$ | $Cu_{11}Sb_3$ |
|---|---|---|---|---|---|
| **Structural data** | | | | | |
| Pearson symbol | *hP*98 | *oC*80 | *oP*8 | *hP*26 | *oC*28 |
| Space group | *P6₃/mmc* | *Cmcm* | *Pmmn* | *P6₃/m* | *Amm*2 |
| Lattice parameters (Å) | *a* = 20.433　*c* = 4.818 | *a* = 5.529　*b* = 47.756　*c* = 4.323 | *a* = 4.230　*b* = 5.280　*c* = 4.540 | *a* = 9.920　*c* = 4.320 | *a* = 4.324　*b* = 19.080　*c* = 4.724 |
| Same phases | $Au_{11}In_3$　$Au_{77}Mg_{23}$　$Cu_{78}Sb_{22}$ | - | $Ag_3Sn$　$Cu_3Sb$　$Au_3In$　$Ag_3Sb$　$Ag_3Bi$ (HP) | $Au_{10}In_3$ | $LiZn_4$ |
| **FS – BZ data from the BRIZ program** | | | | | |
| z (number of valence electrons per atom) | 1.33 | 1.75 | 1.75 | 1.7 | 1.75 |
| $k_F$ (Å$^{-1}$) | 1.304 | 1.537 | 1.601 | 1.526 | 1.550 |
| Total number BZ planes | 26 | 34 | 22 | 44 | 34 |
| $k_F/(½q_{hkl})$　max | 1.060 | 1.057 | 1.078 | 1.049 | 1.067 |
| min | 0.929 | 1.021 | 0.977 | 0.962 | 0.965 |
| $V_{FS}/V_{BZ}$ | 0.804 | 0.903 | 0.912 | 0.946 | 0.898 |

Notes: structural data are from [18] and from papers cited in the related paragraphs of the text.

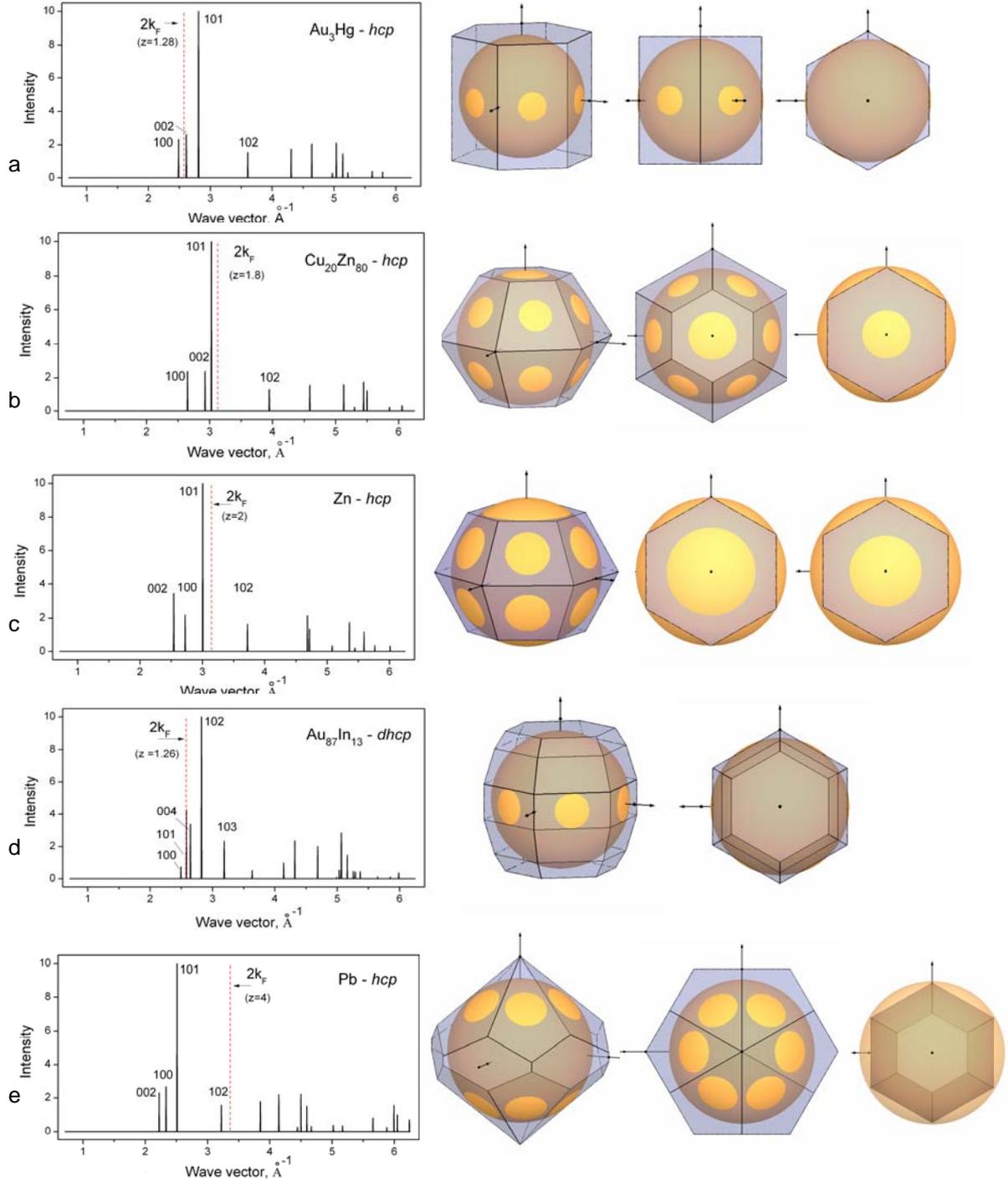

**Figure 1.** Calculated diffraction patterns for selected phases from Table 1 (left panels) and corresponding Brillouin-Jones zones with the inscribed Fermi spheres (right panels). The position of $2k_F$ and the hkl indices of the planes used for the BZ construction are indicated on the diffraction patterns. The FSBZ configurations are shown with the common view (first left) and down c* (middle and right). (c) For *hcp*-Zn projections down c* are shown for the real $c/a$ = 1.856 (middle) and for the ideal $c/a$ = 1.633 (right). (e) For *hcp*-Pb projections down c* are shown for polyhedra formed by (102) and (110) planes (middle) and by (101) and (002) planes (right).

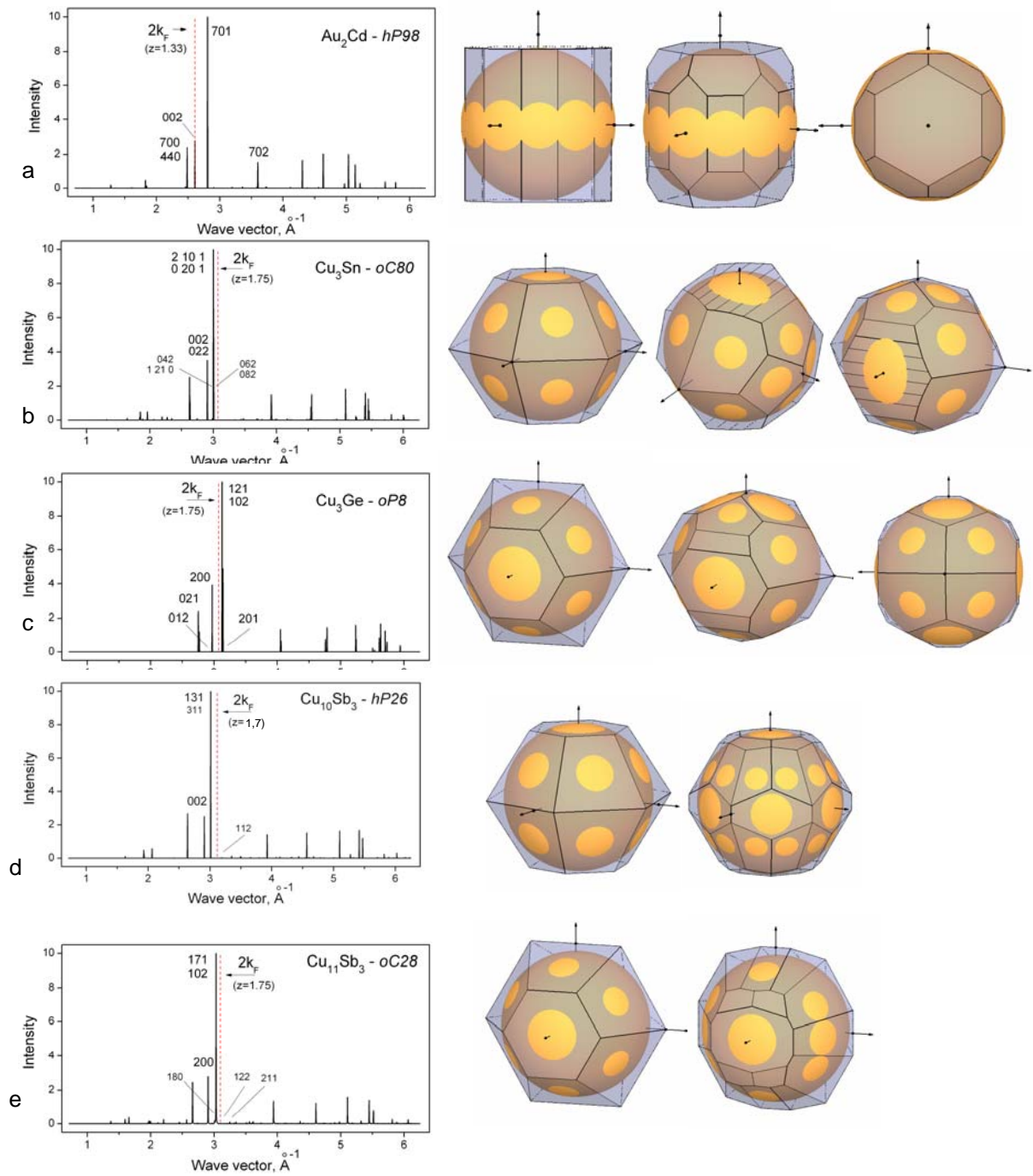

**Figure 2.** Calculated diffraction patterns for selected phases from Table 2 (left) and corresponding Brillouin-Jones zones with the inscribed Fermi spheres (right). The position of $2k_F$ and the hkl indices of the planes used for the BZ construction are indicated on the diffraction patterns.